\documentclass[acmsmall]{acmart}

\AtBeginDocument{%
  }

\usepackage{subcaption}
\usepackage{dcolumn}
\newcolumntype{d}[1]{D..{#1}}

\begin{document}

%%
%% The "title" command has an optional parameter,
%% allowing the author to define a "short title" to be used in page headers.
%\title{The Spectator Perspective from the Spectator Perspective}
%hide and seek
%\title{What I Show and What You See: Studying The Witness Experience
%from the Witness Perspective}

\title{Beyond Hiding and Revealing: Exploring Effects of Visibility and Form of Interaction on the Witness
Experience}
%Interaction Design and Perceived Form on User Interactions and Witness
%Perception}

%\title{Beyond the User: Investigating the Role of Interaction Design on Witness
%Experiences in Human-Computer Interaction}

%%
%% The "author" command and its associated commands are used to define
%% the authors and their affiliations.
%% Of note is the shared affiliation of the first two authors, and the
%% "authornote" and "authornotemark" commands
%% used to denote shared contribution to the research.

\author{Alarith Uhde}
\email{alarith.uhde@uni-siegen.de}
\orcid{0000-0003-3877-5453}
\affiliation{%
  \institution{University of Siegen}
  \streetaddress{Kohlbettstraße 15}
  \city{Siegen}
  \country{Germany}
  \postcode{57072}
}
\affiliation{%
  \institution{Tokyo College, The University of Tokyo}
  \country{Japan}
}

\author{Tim zum Hoff}
\email{tim.zumhoff@uni-siegen.de}
\orcid{0000-0002-1785-8228}
\affiliation{%
  \institution{University of Siegen}
  \streetaddress{Kohlbettstraße 15}
  \city{Siegen}
  \country{Germany}
  \postcode{57072}
}

\author{Marc Hassenzahl}
\email{marc.hassenzahl@uni-siegen.de}
\orcid{0000-0001-9798-1762}
\affiliation{%
  \institution{University of Siegen}
  \streetaddress{Kohlbettstraße 15}
  \city{Siegen}
  \country{Germany}
  \postcode{57072}
}

%%
%% By default, the full list of authors will be used in the page
%% headers. Often, this list is too long, and will overlap
%% other information printed in the page headers. This command allows
%% the author to define a more concise list
%% of authors' names for this purpose.
%\renewcommand{\shortauthors}{Uhde et al.}

%%
%% The abstract is a short summary of the work to be presented in the
%% article.
\begin{abstract}

  Our interactions with technology do not just shape our individual experiences.
  They also affect people around us. Although previous research has addressed
  such ``witness'' experiences, the actual effect of interaction design on the
  witness experience remains largely unknown. In an online study (n = 407), we
  explored how witnesses perceive mid-air gesture-based interactions with
  a hearing aid, using four video vignettes. We studied witnesses' subjective
  visibility of manipulations and effects (following Reeves and colleagues'
  taxonomy), perceived form of interaction, subjective experience, and
  relationships between these measures. Although visibility patterns matched the
  intended form, they did not lead to the supposed experience (i.e.,
  ``suspenseful'' gestures did not lead to suspenseful experiences). The paper
  illustrates gaps in current research about witness experiences, demonstrates
  the need to overcome basic hiding/revealing profiles, and indicates a path
  forward by focusing on aesthetic forms and experiences.

\end{abstract}

%%
%% The code below is generated by the tool at http://dl.acm.org/ccs.cfm.
%% Please copy and paste the code instead of the example below.
%%



\setcopyright{acmlicensed}
\acmJournal{PACMHCI}
\acmYear{2023} \acmVolume{7} \acmNumber{MHCI} \acmArticle{200} \acmMonth{9}
\acmPrice{15.00}\acmDOI{10.1145/3604247}

%%
%% Keywords. The author(s) should pick words that accurately describe
%% the work being presented. Separate the keywords with commas.

\received{January 2023}
\received[revised]{May 2023}
\received[accepted]{June 2023}

%%
%% This command processes the author and affiliation and title
%% information and builds the first part of the formatted document.
\maketitle

\section{Introduction}

Guitar shops around the world have an unspoken rule. Customers are free to try
out all the instruments and play everything they feel confident enough to show
off in front of others. There is one exception, though: ``Stairway to Heaven'',
or as they also call it: ``the forbidden riff''. The reason is that this song is
relatively easy and fun to play, so every day dozens of customers will come in
and play it, over and over again, to the point that it becomes unbearable for
the shop owners. The stairway problem is so pervasive that it even made it into
a Hollywood movie: ``No Stairway!
Denied!''\footnote{\url{https://www.youtube.com/watch?v=RD1KqbDdmuE}}. Thus, if
you want to buy a guitar and absolutely need to play Stairway to Heaven to make
your shopping decision, maybe ask for headphones so that the shop owner only
sees you play, but does not need to hear that song again.

In guitar shops, we can think of this use of headphones as a simple way to make
the life easier for the shop owners, or in other words to ``improve their
witness\footnote{Note that previous terms used in the literature to refer to
``other people'' around the user include ``spectators'' \citep{Reeves2005},
``audience'' \citep{Rico2010}, ``attendant'' \citep{VonTerzi2023}, or
``observers'' \citep{Alallah2018}. We use ``witness'' here to highlight that the
other people do not always deliberately focus on the user.} experience''.
Headphones specifically hide the ``effects'' of the guitar interaction (the
music), while keeping the ``manipulations'' (how the user places their fingers)
revealed as they were. This seems appropriate in the particular situation of
a guitar shop, because it specifically removes the source of irritation. Perhaps
surprisingly, then, we find that the Human-Computer Interaction (HCI) literature
explicitly advises against this.  Such design interventions that keep
manipulations revealed and hide the effects are commonly considered socially
unacceptable (e.g., \citep{Ens2015, Monk2004a, Montero2010, Koelle2020}).
Instead, current recommendations to improve the witness experience would be to
either reveal the effects (e.g., unplug the headphones again, so the shop owner
can hear the music; \citep{Ens2015}), or to hide the manipulation along with
them \citep{Rico2010, Koelle2020}.

Why is there such a stark contrast between the world of guitar shops and current
design recommendations in HCI? One reason could be that the mechanisms of how
interactions with technology influence witness experiences are still not well
understood. The major breakthrough in this area was a taxonomy by Reeves and
colleagues \citep{Reeves2005}, published eighteen years ago. They distinguished
whether an interaction hides or reveals a) manipulations and b) effects, as in
the example above, and mapped this to certain witness experiences. This caused
a series of follow-up studies, each comparing interactions of the different
categories and how people experience them (e.g.,~\citep{Ahlstrom2014,
Alallah2018, Montero2010, Ens2015, Uhde2022d}). However, despite its wide
adoption, there is surprisingly limited follow-up work that analyzed the
\emph{in-between} in more detail: Everything that happens between the
interaction design process, where designers think about hiding and revealing
certain parts of an interaction, to the actual interaction as the user performs
it, to the perceived interaction from the witness perspective, and finally their
situated experience.  In addition, we argue that the taxonomy has led to
a relatively narrow focus on whether a certain part of an interaction should be
hidden or revealed, and away from \emph{how} this hiding and revealing should
look like. Finally, the taxonomy is to a large extent designer- and
user-centered, and not witness-centered. This has led to some unintuitive and
ambiguous definitions of core concepts, as we outline below.

In this paper, we take a first step to further explore the relationship between
perceived attributes of an interaction and the witness experience.
Specifically, we analyze whether witnesses in fact perceive the revealed and
hidden parts of a user's interaction as intended by the designer, and whether
this leads to experiences as suggested by the taxonomy. Unlike previous work
that broadly covered variations of hidden or revealed manipulations and effects
to compare them (e.g., \citep{Ens2015, Montero2010, Koelle2020}), we
specifically focused on one category of interactions only: ``Suspenseful''
interactions, with revealed manipulations and hidden effects.  This is the most
controversial category in the literature. It has led to a wide range of user and
witness experiences in the past, including positive \citep{Reeves2005,
Uhde2022d} and (mostly) negative examples \citep{Ens2015, Koelle2020,
Montero2010}. We took this variety as an opportunity to study more fine-grained
differences in the forms of interaction and their effects on witness
experiences.

\section{Background}

Research on witness experiences became more important and common in the wake of
the ``third wave of HCI'' \citep{Bodker2006}, when new interactive technologies
flooded social spaces. At the time, researchers increasingly focused on
technology-mediated experiences in social situations (e.g., \citep{Ahlstrom2014,
Reeves2005, Rico2010, Koelle2020, Uhde2021b, Uhde2022d}), often driven by the
rise of the smartphone \citep{Love2004, Monk2004a, Monk2004b, Okabe2005,
Turner2008} and the introduction of other mobile technologies such as smart
glasses \citep{Alallah2018, Hakkila2015, Koelle2015, Koelle2017, Koelle2018},
fitness trackers \citep{Hassenzahl2016b, Karapanos2016}, and wearables
\citep{Dagan2019, Epp2019, Kelly2016, Muehlhaus2022, Profita2013}. A major part
of this research investigated design strategies to make technologies more
``socially acceptable'' \citep{Kelly2016, Koelle2017, Koelle2020, Montero2010,
Rekimoto2001, Uhde2021b}. These studies typically focus on avoiding negative
experiences. Common strategies include making technologies less obtrusive by
hiding them in accessories \citep{Rekimoto2001}, or making them more ``subtle''
\citep{Rico2010}. Another approach is to clearly communicate the purpose of an
interaction to potential witnesses to avoid misunderstandings
\citep{Koelle2020}. In addition to social acceptability, some researchers have
also looked into ways in which technologies can be used to create or facilitate
positive experiences, often as tools to initiate social interactions
\citep{Mitchell2017, Mitchell2019, Olsson2020, VonTerzi2021}.

When studying how people experience interactive technologies in social
situations, a major complication is that there are various people involved, who
all may experience the interactions differently. Previous research has mainly
focused on the users' immediate experience of their own interaction
\citep{Hassenzahl2006, Hassenzahl2010a}. But in social situations, we can also
consider the witness experience of the users' interactions (e.g.,
\citep{Alallah2018, Baumer2017, Reeves2005}), or even the users' assumptions
about the witness experience and so forth (such potentially endless social
dynamics between co-located people and their assumptions about each other were
analyzed in detail by Goffman \citep{Goffman1959}).

\subsection{Designing for the Witness Experience}

Reeves and colleagues' \citep{Reeves2005} taxonomy systematized the relation
between particular properties of an interaction and a witness experience. They
suggested a simple categorization of interaction design patterns, depending on
what the designer hides from or reveals to potential witnesses. According to
this taxonomy, designers can hide or reveal manipulations (e.g., arranging
fingers on a guitar to play a chord) and effects (e.g., the music). This leads
to a 2x2 schema with four resulting design patterns. ``Expressive'' interactions
reveal both manipulations and effects to witnesses. ``Magical'' interactions
hide the manipulations but reveal the effects. ``Suspenseful'' interactions
reveal the manipulations, but hide the effects. And ``secretive'' interactions
hide both the manipulations and effects.

As the names of the four patterns indicate, the authors link each of them with
certain qualities of the witness experience. For example, they associate magical
interactions (hidden manipulations, revealed effects) with a magical experience
(see also \citep{Marshall2010}). Accordingly, if a designer wanted to build
a new ``magical'' guitar, they could try to hide parts of the user's
manipulations but keep the musical effects revealed. So instead of using their
right hand to strum chords which is clearly visible to witnesses, the user could
secretly tap their foot, which may go unnoticed. The four design approaches form
the core of the taxonomy, although the authors also suggest more nuanced
variations. For example, the designer could not only reveal, but even amplify
the effects by adding flashing lights to the guitar, or they could distort the
mapping between manipulations and effects \citep{Reeves2005}.

The original taxonomy was based on example technologies used on stage, during
exhibitions, or in other public performances with an actual ``audience''.
However, it was later applied to a broader scope of social situations. Some
examples include interactions with phones on trains \citep{Monk2004a,
Koelle2020} and in offices \citep{Ens2015}, with vending machines
\citep{Montero2010}, or with a hearing aid in a face-to-face conversation
\citep{Uhde2022d}. All in all, the taxonomy helped to systematize research about
witness experiences across social situations and provided a basic vocabulary.
Nevertheless, several issues became apparent over time that we outline in the
following.

\subsubsection{Three understandings of ``hiding'' and ``revealing''}

One issue has to do with the concepts of hiding and revealing, and what they
mean as design strategies. The original paper gives no explicit definition.
However, based on the examples provided and on later work, we identified three
different ways to think about hiding and revealing in HCI.

The first is to understand hiding as designing something so that the witnesses
cannot perceive it, on a sensory level. In contrast, revealing makes it
perceptible. For example, the manipulation of a smartwatch that measures the
heart rate is hidden in this sense, because even an informed witness cannot know
whether an interaction actually takes place. In contrast, tapping on the
smartwatch is an example for a revealed manipulation. We call this
\emph{perception-based}. For the sake of simplicity, we will adopt this
definition for the rest of this paper unless otherwise noted.

The second way to think about hiding an interaction is to make it go unnoticed
by resembling ``typical'' interactions in a certain situation. In contrast,
revealing lets an interaction stand out from the usual. We call this approach
\emph{convention-based}. Here, witnesses can perceive the interaction on
a sensory level, but it does not draw their attention. This concept is for
example used in ``subtle'' design. A hidden manipulation in this sense could
take the form of foot tapping \citep{Rico2010}, which seems natural and
unremarkable in many situations. Witnesses can perceive it, but they usually do
not know that it is a manipulation of a technology and pay no attention. Perhaps
counter-intuitively, a revealed manipulation in terms of the convention-based
approach can be part of a magical interaction following the perception-based
approach. Think about the ``magical guitar'' example from above. Witnesses of
a guitar concert, uninformed that the musician uses such a magical guitar, would
usually expect to see the fingers on the strings and hear the music. They would
also focus on the playing and the music. But if the musician strums the chords
by tapping their foot, this would break witnesses' expectations and stand out
from the usual. The form of guitar interaction itself becomes a notable focus of
attention and in a way ``revealed'' -- although it is more ``hidden'' according
to the perception-based approach.

The third, \emph{experience-based} approach is to think of ``hiding'' in terms
of the actual witness impression that something is hidden from them, and
``revealing'' as the impression that something is revealed to them. This
definition is somewhat detached from the form of interaction itself, and it
focuses primarily on how the witness experiences it. A ``hidden'' example from
the HCI literature is TongueBoard \citep{Li2019}, an in-mouth device that tracks
the user's tongue movement to recognize speech without the need to actually
utter it. Witnesses can see that the user is speaking, but the entire setup
communicates to them that something is hidden (namely the voice and content of
the conversation). An everyday example is whispering, perhaps performed by
directing the head towards someone else's ear while covering the mouth with
a hand. In contrast, a passenger talking loudly on their phone on a train, to
the point that the other passengers feel that they are addressed in some form as
the ``audience'' of the phone call (see e.g., \citep{Love2004}) could leave the
witnesses with the impression that the user wants to ``reveal'' something to
them.

\subsubsection{Three perspectives on designing the witness experience}

These three understandings of hiding and revealing correspond with three
different approaches we can take when designing for witness experiences. The
perception-based approach most closely matches with Reeves and colleagues' use
of the terms in the original paper \citep{Reeves2005}, and it fits with how
a designer may think about the interaction-experience relation. But it is also
most detached from the actual witness experience. For example, it seems unlikely
that a ``secretive'' interaction with a non-perceptible manipulation and
non-perceptible effects actually creates a secretive experience as suggested.
A witness does not even witness it, and consequently it creates no experience at
all. Similarly, typical ``magical'' interactions, such as a public display in
a train station that is controlled from a remote office also takes the focus
away from the manipulation as such and acts as a pure information output, often
with no magical experience.

The convention-based approach relates to the specific social situation and the
user perspective, who can in some cases decide to hide or reveal parts of the
interaction through their performance. For example, if the technology affords
them to tap their foot, they can do that in a subtle way or draw attention by
tapping with eccentric movements. Here, witnesses can perceive secretive
interactions, but they typically stay ``hidden in plain sight''. As pointed out
above, hidden and revealed interactions according to the perception-based and
the convention-based approach do not always overlap. In some cases, designing
something to be non-perceptible can actually make it stand out. Thus,
``revealing'' might better be thought of as ``drawing attention to'' something
by deviating from conventions. This can make users feel uncomfortable in some
situations \citep{Koelle2020}, and has been suggested as a reason why
suspenseful interactions were described as ``awkward'' in the past
\citep{Ens2015, Montero2010}. A further complication with the convention-based
approach is that interactions considered ``normal'' today may vary over time or
with different user/witness demographics (e.g., \citep{Okabe2005, Shove2012}).

Finally, the experience-based approach does not take the detour through
perceptibility or ``normal'' behavior, and instead directly focuses on the
actual experience of the witness. One problem for interaction designers is that
this does not correspond to a simple set of rules about how the interaction
should look like. There are certain established practices that can create such
experiences, but they are not easily applicable for all technologies. In
addition, the point of the taxonomy was to match characteristics of an
interaction with experiential consequences, and not experiences with other
experiences.

\subsection{Witness Experiences and the Form of Interaction}

It seems that the relationship between the form of interaction and the witness
experience is more complicated. Not only do different patterns of hiding and
revealing lead to different experiences. Even the \emph{same} form of
interaction can lead to different experiences. Uhde and colleagues
\citep{Uhde2021b} described how experiences of a standard phone call can vary.
When performed in a library, witnesses may feel disturbed. On a rock concert or
in a shopping street, witnesses may not bother. On a train, some curious
witnesses might enjoy listening in \citep{Desmet2012, Turner2008}. They explain
these differences not only based on the user's interaction, but also on what the
witness is doing and other situational characteristics. A person in a library
experiences a phone call negatively because they are trying to read, and the
phone call interferes with this. In other words, some factors that influence the
witness experience can be outside the interaction designer's control.

As a final remark, we want to note that witness experiences \emph{can} be more
predictable, when focusing on specific cases. For example, the taxonomy
initially considered stage performances. Here, spatial position, social
conventions, and typical activities of the witnesses are mostly set. But in more
chaotic or varied social situations, a lot of unexpected things can happen.
Perceptive barriers like a train seat or a window can hide parts of an
interaction from the witnesses' eyes and ears. Different witnesses can consider
different behaviors as ``conventional'', and thus experience the
hiding/revealing profile of the same interaction differently (in terms of the
convention-based approach). They may also perform various activities, from
sleeping, to working, playing games, or listening closely to the user, all of
which can have their own experiential consequences.

Our focus so far was on hiding and revealing, mainly to clarify the importance
of distinguishing between the different perspectives on designing for the
witness experience. However, ``hidden'' or ``revealed'' are relatively vague
descriptions of an interaction, and they are often used as binary categories
(e.g., \citep{Ens2015, Montero2010, Koelle2020, Uhde2022d}). This categorization
can be useful, but it also drastically reduces the design space.  In the social
acceptability literature that we have mostly covered so far, the actual form of
manipulations and effects has only played a subordinate role, compared with the
decision to hide or reveal them.

Other parts of the HCI literature have analyzed the form of interaction in more
detail, although mostly in terms of the user experience (e.g.,
\citep{Diefenbach2017c, Lenz2013}). Of course, the same restrictions about the
impact of witness activities and other situational differences apply here as
well. Nevertheless, they found some links between certain characteristics of an
interaction and its experiential qualities. In a case study on keeping ``small
secrets'' with an interactive photo frame that can hide a picture, participants
found slow, gentle, direct, and instant forms of interaction suitable
\citep{Diefenbach2017c, Lenz2013}. In contrast, they described the suitable form
of revealing an engagement ring from its case as slow, stepwise, delayed,
precise, gentle, and targeted \citep{Lenz2013}. These descriptions of suitable
forms of interaction can also inform more fine-grained sub-categories, within
the same class of interaction. A slow, fluent, delayed, and gentle corkscrew
interaction fits with an experience of relatedness, for example during a picnic
\citep{Lenz2017}. In contrast, a fast, stepwise, instant, and powerful corkscrew
interaction fits with an experience of competence -- suitable for professionals
in an expensive restaurant. When designing for witnesses, such detailed profiles
of forms of interactions and their experiential effects could extend the current
focus on hiding and revealing.

\subsection{Summary and Research Questions}

All in all, the relationship between hiding and revealing parts of an
interaction and the witness experience seems more complex than currently framed
in the literature. We outlined different perspectives of the designer, user, and
witness, each of which highlights and potentially overlooks different parts of
the process. Previous research has mostly focused on the designer's side
(hiding/revealing patterns), rather than studying how the witnesses actually
perceive and experience the interaction and its form in greater detail. To
complement this research, the following study focuses mainly on the witness.

The main goal of our study was to shed some light on the witness perspective and
on how they experience different forms of interaction. We focused on suspenseful
interactions only, because this category has led to the broadest variety of
experiences in the past and these experiences were our central interest. Keeping
the category constant allowed us to study finer differences within it. In sum,
our research questions were:

\begin{description}
  \item[RQ1:] How do suspenseful design patterns (revealed manipulations
    and hidden effects) correspond to subjective visibility from the witness perspective?
  \item[RQ2:] How do witnesses perceive the form of interaction of different
    suspenseful gestures?
  \item[RQ3:] How do witnesses experience suspenseful gestures in terms of the four
    experiential qualities suggested by the taxonomy (i.e., secretive, suspenseful, expressive, and magical)?
  \item[RQ4:] How do the visibility (RQ1) and form of interaction (RQ2) shape
    the witness experience (RQ3)?
\end{description}

\section{Method}

We ran an online study on SurveyMonkey\footnote{https://www.surveymonkey.com}
and recruited participants through
Clickworker\footnote{https://www.clickworker.com}, where they received a link to
access our study. As a test scenario, we chose a conversation between two people
during a cocktail party. This scenario offers a typical social situation where
people interact with strangers, so closely watching someone interacting with an
unfamiliar technology seemed relatively natural. It also provided a stable
setting that we could stage realistically, with limited visual distraction
(e.g., no need to move around, no need to focus on a stage), and that seemed
easy for a broad range of participants to immerse into. As an interactive
technology, we chose a futuristic, interactive hearing aid worn by the
conversation partner. This technology directly relates to the broader
interaction (a conversation), so there was no need to introduce an additional,
unrelated activity. However, it is sufficiently unfamiliar to avoid confounding
effects based on pre-existing manipulation-effect mappings. Finally, it
naturally affords suspenseful forms of interaction.

\subsection{Participants}

We recruited 414 German-speaking participants in total, 7 of which were excluded
from the analysis because of technical issues (details below). Out of the
remaining 407 participants, 241 identified as male, 160 as female, 2 as diverse,
and 4 did not reply. The average age was 37.50 years ($sd = 12.00$; $min = 18$;
$max = 70$).

The most frequent occupation among the participants was ``student'' ($n=49$),
followed by freelancers ($n=17$), engineers ($n=15$), clerks ($n=14$),
unspecified jobs in IT ($n=14$), and 14 missing responses. There was also a long
tail of various jobs with five or fewer mentions ($n=206$ participants in
total).

\subsection{Procedure}

\subsubsection{Introduction and technical checks}

On the first page, we included general instructions. We informed participants
about the anonymous data collection and the expected length of the study (10
minutes). We also asked them to check that their audio setup is working
correctly. As an additional check, we played back a short video on the next
page, in which a number was said that we asked the participants to select from
a drop-down menu. Two participants selected the wrong number and were excluded
from further analysis.

\subsubsection{Four variations of suspenseful interactions}

\begin{figure}[b]
\centering
\includegraphics[width=0.9\linewidth]{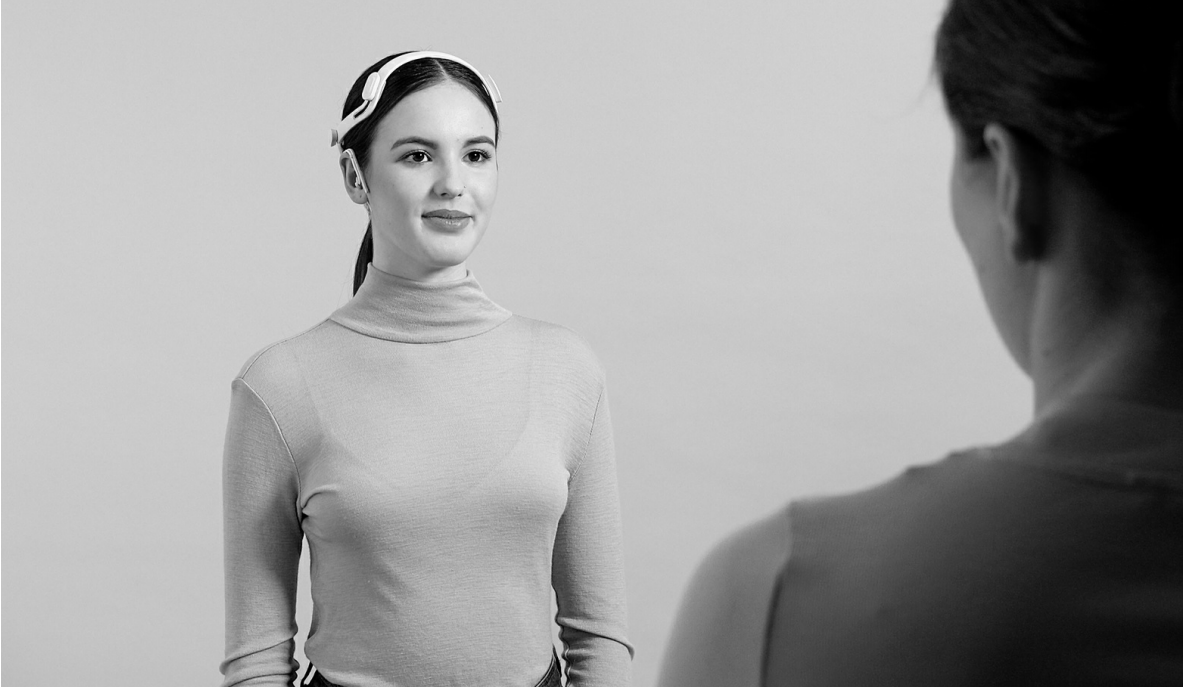}
\caption{The screenshot of the videos used to describe the setting to
  participants.}%
\Description{Two people in a conversation. Detailed description in the main text.}%
\label{fig:video-screenshot}
\end{figure}

On the next page, we showed participants a screenshot from a short video scene
they were about to see (Figure~\ref{fig:video-screenshot}). The image shows two
people facing each other during a conversation. The person on the left side of
the image (the user) is clearly visible from the front. She is wearing
a futuristic hearing aid that spans around her head. The person on the right
(the witness) is slightly blurred and visible from the back, and we asked
participants to imagine being that person during this conversation. Such
over-the-shoulder shots are commonly used in film making to frame a situation
from the perspective of one person, and we used it here to reinforce the
perspective of the witness. We asked participants to imagine talking about their
hobbies. At one point during the conversation, there would be a disturbance and
a reaction to that disturbance, which they would see in the video.

The interaction scenario was based on a prototypical ``beam forming'' function
of the hearing aid, developed in the context of a larger research project. The
hearing aid can ``focus'' on a sound source, such as the conversation partner,
and fade out surrounding noise. But this function does not always work
seamlessly. Sometimes the focus can get distracted, for example when there is
a loud and sudden, external noise. In such a case, the conversation partner's
voice is accidentally faded out and the noise is amplified. The user can
manually readjust the focus to the conversation partner, for example through
a hand gesture. However, note that we did not inform participants about the
function of the hearing aid and the meaning of the interaction at this point.

\begin{figure*}[b] \centering \begin{subfigure}{0.49\textwidth}
\includegraphics[width=0.9\textwidth]{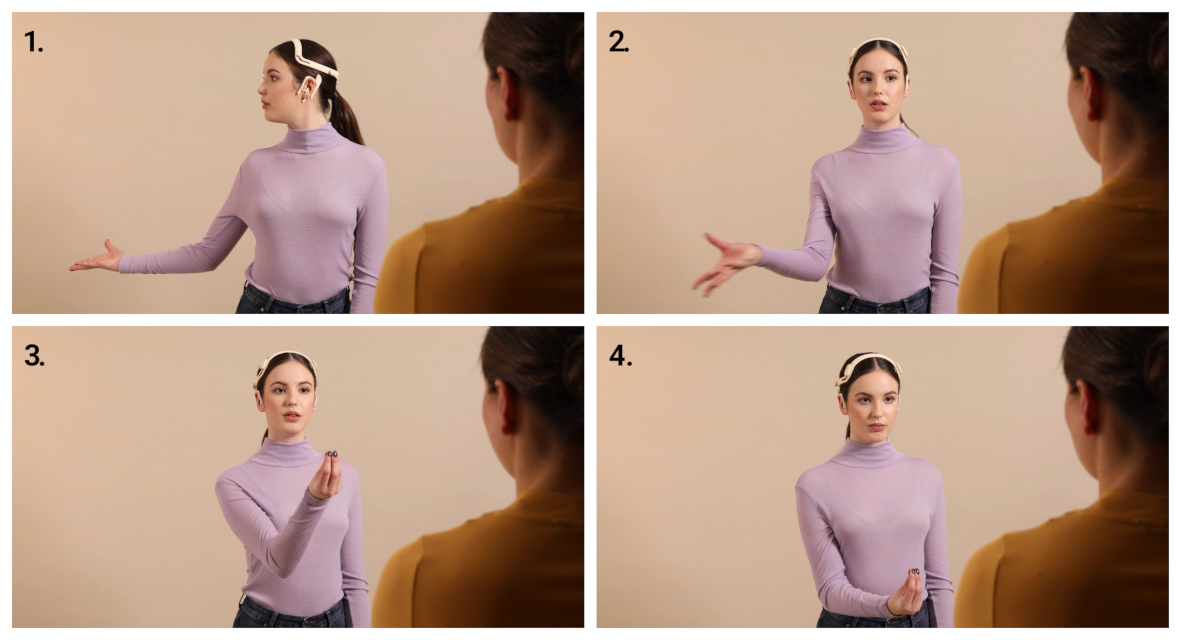} \caption{``settle''}
\end{subfigure} \begin{subfigure}{0.49\textwidth}
  \includegraphics[width=0.9\textwidth]{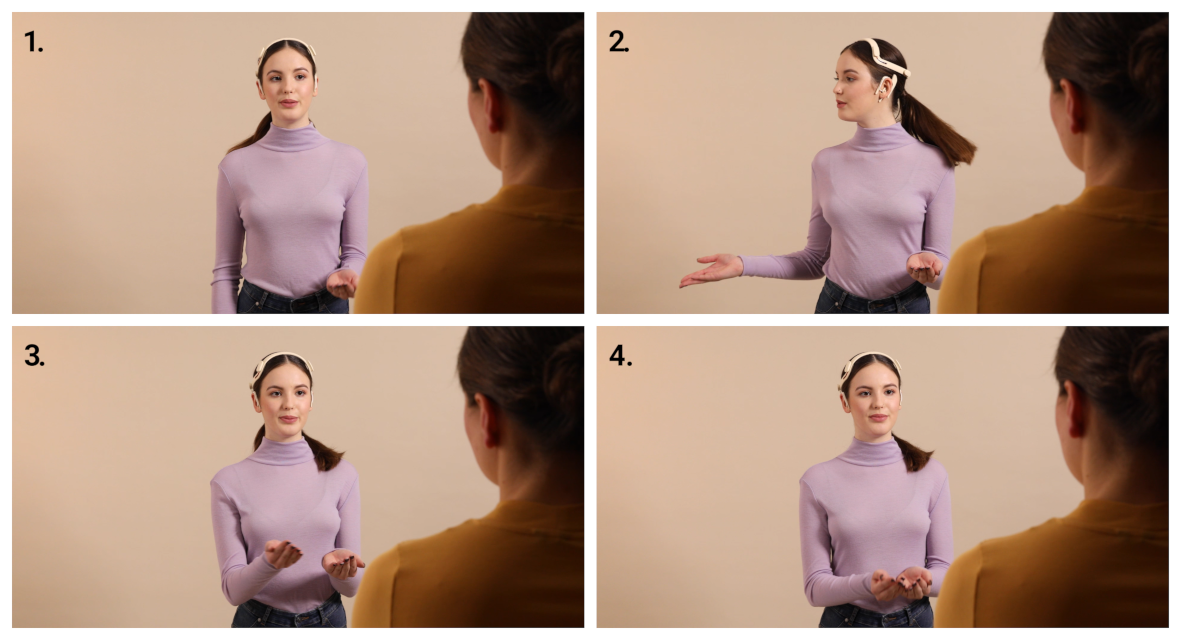} \caption{``across''}
\end{subfigure} \begin{subfigure}{0.49\textwidth} %\centering
\includegraphics[width=0.9\textwidth]{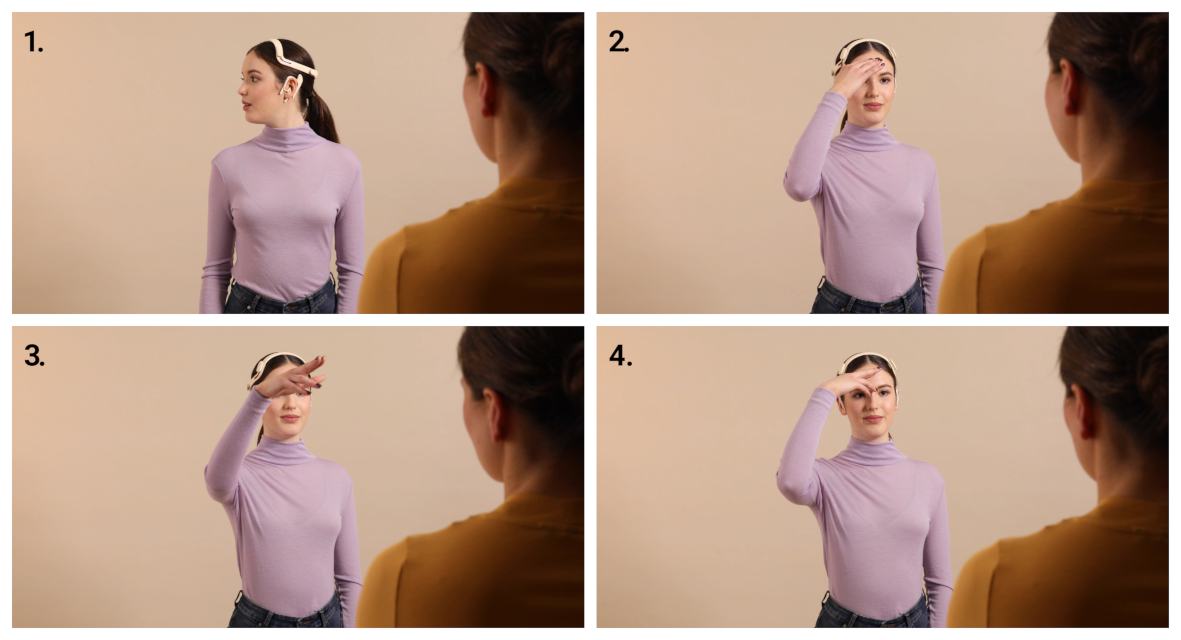} \caption{``grasp''}
\end{subfigure} \begin{subfigure}{0.49\textwidth} %\centering
\includegraphics[width=0.9\textwidth]{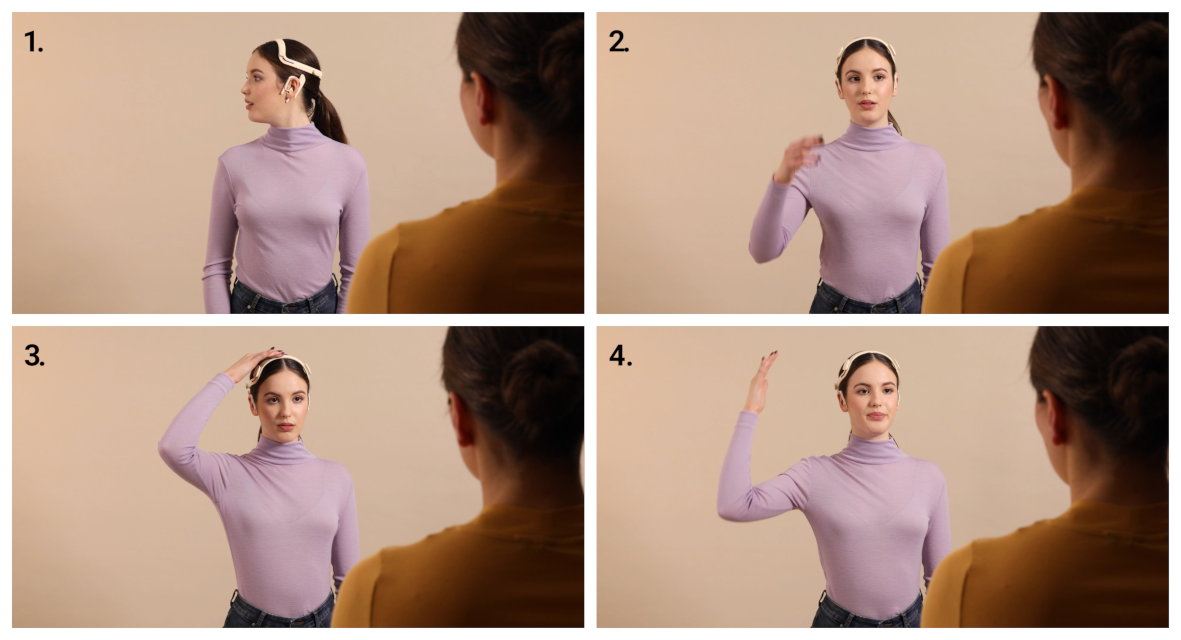} \caption{``tap''}
\end{subfigure} \caption{The four interactions. ``Settle'' and ``across'' (top)
make a spatial reference to the source of distraction. ``Settle'' and ``grasp''
(left) make a spatial reference to the conversation partner.}%
\label{fig:four-interactions} \Description{Four forms of interaction with the
hearing aid, each depicted in a series of four photos. The first interaction
(settle) begins by reaching out to the source of distraction, with an open hand
(the inner side of the hand shows upwards). The user then symbolically ``grabs''
the hearing focus, moves the hand with the ``grabbed'' focus to the desired
direction, and opens it again, thus ``releasing'' the audio focus. The second
interaction (across) begins by holding an open hand in front of the user's body,
close to the hip and subtly pointing to the desired direction. Then the user
opens the other hand and points it to the distracting noise and brings it back
to the front of her body, next to the other hand. The third interaction (grasp)
begins by only looking at the distraction, then looking back at the conversation
partner. The user than uses two fingers to point at the front of her own head,
then to the conversation partner, and finally back to her own head. The fourth
interaction (tap) also begins by only looking at the distraction source and then
back to the conversation partner. The user then taps on the hearing aid for
a moment, on the top of her head, while looking at the conversation partner. She
then releases the hearing aid again.}
\end{figure*}

Participants were randomly assigned to one out of four experimental conditions.
In each condition, they saw a different hand gesture to interact with the
hearing aid in reaction to the disturbance (Figure~\ref{fig:four-interactions}).
We chose a between-subjects design, because we wanted to avoid repetition
effects that could be critical when testing unfamiliar interactions, and because
we had access to a sufficiently large sample. The specific interactions stemmed
from a set of 28 gestures developed in a workshop with a professional actor in
the context of the research project. During this workshop, we introduced the
actor to eight interaction scenarios in social situations, such as ``having
brief discussions with a group of people'' or ``walking down a busy street while
talking and monitoring cars''. We went through different purposes of interaction
in these scenarios, such as ``change focus from person 1 to person 2 manually''.
The actor then tried to immerse into the situation and to develop aesthetic,
meaningful, and discernible gestures (resembling a Bodystorming approach
\citep{Oulasvirta2003}).

This set offered a broad range of gestures from which we selected four, with the
goal to represent a large variety of suspenseful interactions. In addition, none
of them directly corresponded to an established gesture that is already
associated with a certain meaning.

All four gestures consist of a clearly revealed hand gesture, but hidden effects
(in fact, the prototype was non-functional). We chose two gestures (``settle''
and ``across'') that make a spatial reference to the source of disturbance, to
which the hearing focus was accidentally distracted. The user reaches out with
their hand towards the sound source to symbolize ``taking the hearing focus
away'' from there. In addition, we selected two gestures (``settle'' and
``grasp'') that make a spatial reference to the conversation partner by moving
the hand towards them, to symbolize ``setting the hearing focus back to the
partner''. The fourth gesture (``tap'') makes no spatial references, and
``settle'' makes references to both the source and the conversation partner.
This combination allowed us to explore potential effects of meaningful
references (from the designer's perspective) ``to the source'' and ``to the
partner'' on the witness experience, independent of each other.

On the next page, participants watched the video, which was 9 to 10 seconds long
and included the form of interaction of their condition. They could repeat it
until they felt they understood what is happening. We also included a checkbox
to confirm that they have watched the video fully, and three participants who
did not check it were excluded from the study. Two further participants did not
continue the study after watching the video and checking the box, and we also
excluded them from further analysis.

The videos contained a short excerpt of the conversation (see the supplementary
material for the full videos). The conversation partner on the right (i.e., the
participant) is talking with an unintelligible, mumbled voice, and the person on
the left is listening affirmatively. We also included lively background noises
of people talking, to create an atmosphere of a cocktail party. After four
seconds, there is a sound of shattering glass coming from the left, and the
conversation partner in the video turns her head towards the sound source
because her focus of hearing is distracted. By using the hand gesture, she
manually resets the focus of her hearing aid back to the conversation partner
(the participant).

After watching this video and possibly repeating it a few times, the
participants moved on to fill in the rest of the questionnaire.

\subsection{Measures}

\subsubsection{Manipulation and effect visibility}

First, we included two items to check whether participants in fact perceived the
visibility pattern of the interaction as ``suspenseful'', according to the
taxonomy. We selected suspenseful interactions only, so we expected all
interactions to have a high visibility of the manipulation and a low visibility
of the effects from the witness perspective. The first item measured the
witnesses' subjective manipulation visibility (``How did you perceive the hand
gesture?'') on a scale from 1 (``not at all visible'') to 4 (``particularly
visible''). The second item measured the witnesses' subjective effect visibility
(``How did you perceive the change in the situation caused by this hand
gesture?'') with the same scale.

\subsubsection{Interaction Vocabulary}

In addition to subjective visibility, we were interested in how participants
perceived the form of interaction itself. To measure this, we included the
Interaction Vocabulary \citep{Diefenbach2013, Lenz2013}, which contains eleven
7-point semantic differentials that measure aesthetic attributes of the
interaction. The eleven items of the Interaction Vocabulary do not imply
a ``better'' or ``worse'' pole, but are constructed as neutral descriptions of
the form. They are: ``slow -- fast'', ``stepwise -- fluent'', ``instant --
delayed'', ``uniform -- diverging'', ``constant -- inconstant'', ``mediated --
direct'', ``spatial separation -- spatial proximity'', ``approximate --
precise'', ``gentle -- powerful'', ``incidental -- targeted'', ``apparent --
covered''.

The items of the Interaction Vocabulary have no inherent factor structure,
because the perception of the form of interaction depends on the concrete
interaction in focus. Correlations among the items can vary across contexts.
Nevertheless, our study covered a specific set of interactions in a predefined
scenario (a conversation during a cocktail party) and focused on their
differences and similarities. Thus, we grouped the items for a comparative
analysis using a Principal Component Analysis (PCA).

We used parallel analysis to extract components and an oblique rotation
procedure (direct oblimin), because correlations between the components seemed
plausible. In a first run, this led to a three component solution. The item
``spatial separation -- spatial proximity'' had low loadings ($<.3$) on all
components, so we excluded it. The final solution based on the other 10 items
consisted of three components and explained 61\% of the total variance (see
Table~\ref{tab:iv-pca}). The first component consisted of the items (positive
poles in the sense of the component only): ``diverging'', ``inconstant'', and
``stepwise'', and we labeled it as ``undetermined''. The second component
consisted of the items ``fast'', ``powerful'', ``instant'', and ``direct'', and
we labeled it as ``energetic''.  The third component included the items
``apparent'', ``targeted'', and ``precise'', and we labeled it as
``straightforward''. We calculated individual scores for each participant on the
three components as the means of the associated items.

\begin{table}[b]
	\centering
%\addtolength{\tabcolsep}{-0.33em}
	\caption{Component Loadings and Eigenvalues of the three Interaction
    Vocabulary components. Only loadings >.4 are shown.}%
    \small%
\label{tab:iv-pca}
	{
		\begin{tabular}{lrrrr}
			\toprule
			item & undetermined & energetic & straightforward & uniqueness  \\
			\cmidrule[0.4pt]{1-5}
			diverging  & $0.86$ & $ $    & $ $    & $0.29$  \\
			inconstant & $0.80$ & $ $    & $ $    & $0.33$  \\
			stepwise   & $0.57$ & $ $    & $ $    & $0.50$  \\
			fast       & $ $    & $0.82$ & $ $    & $0.31$  \\
			powerful   & $ $    & $0.76$ & $ $    & $0.38$  \\
			instant    & $ $    & $0.67$ & $ $    & $0.38$  \\
			direct     & $ $    & $0.45$ & $ $    & $0.58$  \\
			apparent   & $ $    & $ $    & $0.83$ & $0.31$  \\
			targeted   & $ $    & $ $    & $0.81$ & $0.37$  \\
			precise    & $ $    & $ $    & $0.52$ & $0.45$  \\
            \addlinespace
            Eigenvalue & $3.20$ & $1.62$ & $1.29$ \\
			\bottomrule
			% \addlinespace[1ex]
			% \multicolumn{5}{p{0.5\linewidth}}{\textit{Note.} Applied rotation method is oblimin.} \\
		\end{tabular}
	}
\end{table}

\subsubsection{Experience measures}

On the next page, we measured the witness experience of the interaction. We
included three attributes for each of the four categories and associated
experiences suggested in the taxonomy \citep{Reeves2005}. There is no
established questionnaire to measure these experiences yet, so we used
self-developed items, broadly based on the descriptions in the original paper.
For ``secretive'' experiences, we included the three attributes mysterious,
obscure, and enigmatic. For ``magical'' experiences, we included the attributes
unnatural, magical, and surprising. For ``suspenseful'' experiences, we included
suspenseful, exciting, and fascinating. And for ``expressive'' experiences, we
used the attributes expressive, dynamic, and lively. We asked participants to
rate how they perceived the interaction on each of these twelve items with
a 7-point scale ranging from 1 (``not at all'') to 7 (``extremely'').

Next, we tested the fit of the items to a four component model as implied by the
taxonomy with a Principal Component Analysis. We set the number of components to
four and used an oblique rotation procedure (direct oblimin), because
correlations between the components seemed plausible (e.g., because the
visibility pattern of ``expressive'' and ``suspenseful'' interactions overlap).
The resulting solution for the ``secretive'', ``suspenseful'', and
``expressive'' items was in line with our expectations, and they loaded on three
distinct components. However, the items of the ``magical'' category were
dispersed across the ``suspenseful'', ``hidden'', and a fourth component. One
reason for this inconsistency of the ``magical'' experience could be that it has
no overlap with the suspenseful form that was the basis of all gestures we
tested. The hidden and revealed components of suspenseful and magical
interactions are disjunct, so the magical items may not have been appropriate
for the present study about suspenseful interactions. Thus, we excluded the
``magical'' items and ran another PCA with three preset components and a direct
oblimin rotation. This led to a three components structure that matched the
three remaining experience categories and explained 73\% of the variance (see
Table~\ref{tab:experience-pca}). We calculated an individual score for each
participant on the three components based on the mean of the associated items.

\begin{table}[t]
	\centering
%\addtolength{\tabcolsep}{-0.33em}
	\caption{Loadings and Eigenvalues of the witness experience components. Only
    loadings >.4 are shown.}%
    \small%
\label{tab:experience-pca}
	{
		\begin{tabular}{lrrrr}
			\toprule
			item & suspenseful & secretive & expressive & uniqueness  \\
			\cmidrule[0.4pt]{1-5}
			exciting    & $0.92$ & $ $    & $ $    & $0.21$  \\
			fascinating & $0.87$ & $ $    & $ $    & $0.23$  \\
			suspenseful & $0.82$ & $ $    & $ $    & $0.23$  \\
			enigmatic   & $ $    & $0.89$ & $ $    & $0.20$  \\
			obscure     & $ $    & $0.89$ & $ $    & $0.24$  \\
			mysterious  & $ $    & $0.78$ & $ $    & $0.32$  \\
			dynamic     & $ $    & $ $    & $0.85$ & $0.27$  \\
			lively      & $ $    & $ $    & $0.84$ & $0.31$  \\
			expressive  & $ $    & $ $    & $0.76$ & $0.39$  \\
			\addlinespace
            Eigenvalue  & $3.52$ & $2.19$ & $0.89$ & \\
			\bottomrule
			% \addlinespace[1ex]
			% \multicolumn{5}{p{0.5\linewidth}}{\textit{Note.} Applied rotation method is oblimin.} \\
		\end{tabular}
	}
\end{table}

\subsubsection{Demographic data}

Finally, we asked participants for their gender (male, female, diverse), age,
and occupation.

\section{Results}

Summary statistics of all measures can be found in Table~\ref{tab:summary}, and
correlations in Table~\ref{tab:correlations}.

\subsection{Subjective Visibility}

As indicated above, all four gestures were designed as ``suspenseful'': They had
a revealed manipulation and a hidden effect. In our analysis, we first tested
whether this pattern corresponded with the subjective visibility from the
witness perspective (RQ1). To that end, we first checked whether participants
perceived the gestures, overall, as rather visible or rather not visible by
comparing the subjective visibility scores with the center of the scale (2.5).

\begin{table*}[t]

\caption{\label{tab:summary}Summary statistics of all measures, overall and
  separately for each gesture. m = mean, sd = standard deviation, md = median,
  hist = histogram, IV = interaction vocabulary.}
\scriptsize
\centering
\addtolength{\tabcolsep}{-0.37em}
  \begin{tabular}[t]{lrcclrcclrcclrcclrccl}
\toprule
\multicolumn{1}{c}{ } & \multicolumn{4}{c}{overall} & \multicolumn{4}{c}{across} & \multicolumn{4}{c}{grasp} & \multicolumn{4}{c}{settle} & \multicolumn{4}{c}{tap} \\
  measure & m & sd & md & hist & m & sd & md & hist & m & sd & md & hist
  & m & sd & md & hist & m & sd & md & hist\\
\midrule
  manipulation visibility & 3.43 & 0.60 & 3.00 & \includegraphics[]{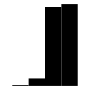}
                          & 3.36 & 0.59 & 3.00 & \includegraphics[]{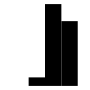}
                          & 3.42 & 0.58 & 3.00 & \includegraphics[]{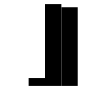}
                          & 3.47 & 0.65 & 4.00 & \includegraphics[]{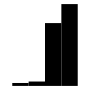}
                          & 3.46 & 0.59 & 4.00 & \includegraphics[]{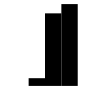}\\
  effect visibility       & 2.55 & 0.97 & 3.00 & \includegraphics[]{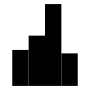}
                          & 2.65 & 0.86 & 3.00 & \includegraphics[]{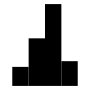}
                          & 2.50 & 1.00 & 3.00 & \includegraphics[]{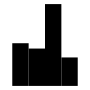}
                          & 2.59 & 1.01 & 3.00 & \includegraphics[]{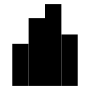}
                          & 2.49 & 0.97 & 3.00 & \includegraphics[]{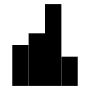}\\

\addlinespace
  IV undetermined         & 3.68 & 0.70 & 3.67 & \includegraphics[]{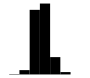}
                          & 3.56 & 0.72 & 3.33 & \includegraphics[]{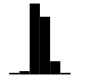}
                          & 3.74 & 0.65 & 3.67 & \includegraphics[]{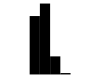}
                          & 3.78 & 0.73 & 3.67 & \includegraphics[]{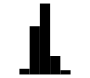}
                          & 3.64 & 0.69 & 3.67 & \includegraphics[]{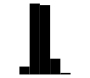}\\
  IV energetic            & 4.14 & 0.70 & 4.00 & \includegraphics[]{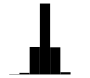}
                          & 4.04 & 0.73 & 4.00 & \includegraphics[]{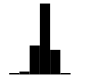}
                          & 4.23 & 0.70 & 4.25 & \includegraphics[]{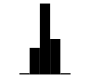}
                          & 4.38 & 0.64 & 4.25 & \includegraphics[]{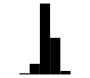}
                          & 3.88 & 0.62 & 4.00 & \includegraphics[]{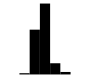}\\
  IV straightforward      & 5.28 & 1.11 & 5.33 & \includegraphics[]{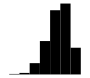}
                          & 5.20 & 1.11 & 5.33 & \includegraphics[]{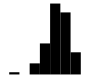}
                          & 5.19 & 1.17 & 5.33 & \includegraphics[]{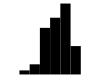}
                          & 5.52 & 1.01 & 5.67 & \includegraphics[]{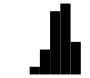}
                          & 5.19 & 1.12 & 5.33 & \includegraphics[]{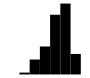}\\
\addlinespace
  secretive               & 4.35 & 1.62 & 4.67 & \includegraphics[]{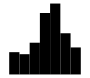}
                          & 4.47 & 1.76 & 4.67 & \includegraphics[]{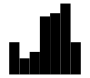}
                          & 4.26 & 1.40 & 4.33 & \includegraphics[]{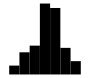}
                          & 4.37 & 1.70 & 4.67 & \includegraphics[]{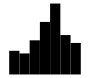}
                          & 4.32 & 1.67 & 4.67 & \includegraphics[]{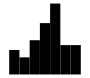}\\
  suspenseful             & 3.44 & 1.40 & 3.67 & \includegraphics[]{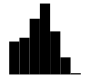}
                          & 3.43 & 1.39 & 3.33 & \includegraphics[]{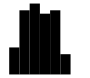}
                          & 3.50 & 1.29 & 3.67 & \includegraphics[]{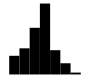}
                          & 3.75 & 1.42 & 4.00 & \includegraphics[]{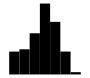}
                          & 3.07 & 1.43 & 3.33 & \includegraphics[]{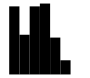}\\
  expressive              & 4.35 & 1.27 & 4.33 & \includegraphics[]{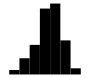}
                          & 4.55 & 1.19 & 4.67 & \includegraphics[]{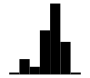}
                          & 4.46 & 1.09 & 4.33 & \includegraphics[]{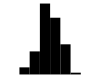}
                          & 4.66 & 1.35 & 4.67 & \includegraphics[]{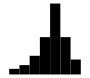}
                          & 3.75 & 1.23 & 4.00 & \includegraphics[]{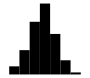}\\
\bottomrule
\end{tabular}
\end{table*}

\subsubsection{Overall visibility fits with suspenseful interaction}

Accordingly, we summarized all scores across the four gestures and ran an
undirected one-sample Wilcoxon test\footnote{We used non-parametric tests for
the visibility scores because we only measured one 4-point item for each of
them.} against the center of the scale. Manipulation visibility was relatively
high (see Table~\ref{tab:summary}), and significantly higher than the center of
the scale ($W = 80531$, $p<.001$, $r = .94$). Conversely, a second undirected
one-sample Wilcoxon test revealed that the effect visibility was not different
from the center of the scale ($W = 44635$, $p=.17$). This is overall consistent
with the visibility pattern of suspenseful interactions, although participants
seemed to have interpreted parts of what they saw as ``effects'' (e.g., the fact
that the conversation could continue after the interaction). Note that the
prototype itself was non-functional, so there was no effect directly related to
the technology which participants could observe.

\subsubsection{No visibility differences between gestures}

In a second step, we tested for differences between the four gestures. We ran
a Kruskal-Wallis tests with gesture type as independent variable and
manipulation visibility as dependent variable. However, there was no significant
effect ($H(3) = 2.82$, $p = .42$). Similarly, we ran a Kruskal-Wallis test with
gesture type as independent variable and effect visibility as dependent
variable, but there was no significant effect ($H(3) = 1.45$, $p = .70$).

Taken together, witnesses perceived the visibility pattern of all gestures
broadly according to the hiding and revealing pattern for suspenseful
interactions. They perceived the manipulation as above-average visible. In
contrast, they perceived the subjective effect visibility as not significantly
different from the center of the scale. In addition, there was no difference in
subjective visibility among the four gestures. In other words, potential
differences in witness experiences between the gestures that we analyze below
are not reflected in differences on the level of pure visibility of
manipulations and effects.

\begin{table}[b]
\addtolength{\tabcolsep}{-0.4em}
	\centering
	\small
	\caption{Pearson correlations between all measures. IV = interaction
    vocabulary. The values in brackets are the internal consistencies (where
    applicable). Correlations marked with an asterisk are significant (*:
    {p<.05}, **: p<.01, ***: p<.001).}%
\label{tab:correlations}
	{
		%\begin{tabular}{lrrrrrrrr}
		\begin{tabular}{@{}ld{1.5}d{1.5}d{1.5}d{1.5}d{1.5}d{1.5}d{1.5}d{1.5}@{}}
			\toprule
			measure & 1 & 2 & 3 & 4 & 5 & 6 & 7 & 8 \\
			\midrule
			1. visibility manipulation &            &           &           &           &          &           &           &       \\
			2. visibility effect       & .19^{***}  &           &           &           &          &           &           &       \\
			3. IV undetermined         & -.02       & .06       & (.71)     &           &          &           &           &       \\
			4. IV energetic            & .12^{*}    & .07       & .15^{**}  & (.67)     &          &           &           &       \\
			5. IV straightforward      & .34^{***}  & .06       & -.13^{**} & .15^{**}  & (.65)    &           &           &       \\
			6. secretive               & .26^{***}  & -.12^{*}  & -.09      & -.02      & .13^{**} & (.82)     &           &       \\
			7. suspenseful             & .07        & .13^{*}   & .04       & .12^{*}   & .02      & .32^{***} & (.85)     &       \\
			8. expressive              & .14^{**}   & .32^{***} & .08       & .33^{***} & .15^{**} & -.00      & .50^{***} & (.76) \\
			\bottomrule
			% \addlinespace[1ex]
			% \multicolumn{15}{p{0.5\linewidth}}{* p $<$ .05, ** p < .01, *** p < .001} \\
			% \multicolumn{15}{p{0.5\linewidth}}{*** $$} \\
			% \multicolumn{15}{p{0.5\linewidth}}{* $$} \\
			% \multicolumn{15}{p{0.5\linewidth}}{** $$} \\
		\end{tabular}
	}
\end{table}

\subsection{Form of Interaction}

In the next step, we analyzed how participants perceived the form of interaction
using the Interaction Vocabulary (RQ2). With this step, we looked for
differences in perception that go beyond pure visibility of the manipulation and
effect.

\subsubsection{All gestures are ``energetic'', ``straightforward'', and not
``undetermined''}

\begin{figure}[b]
\centering
\includegraphics[width=0.9\linewidth]{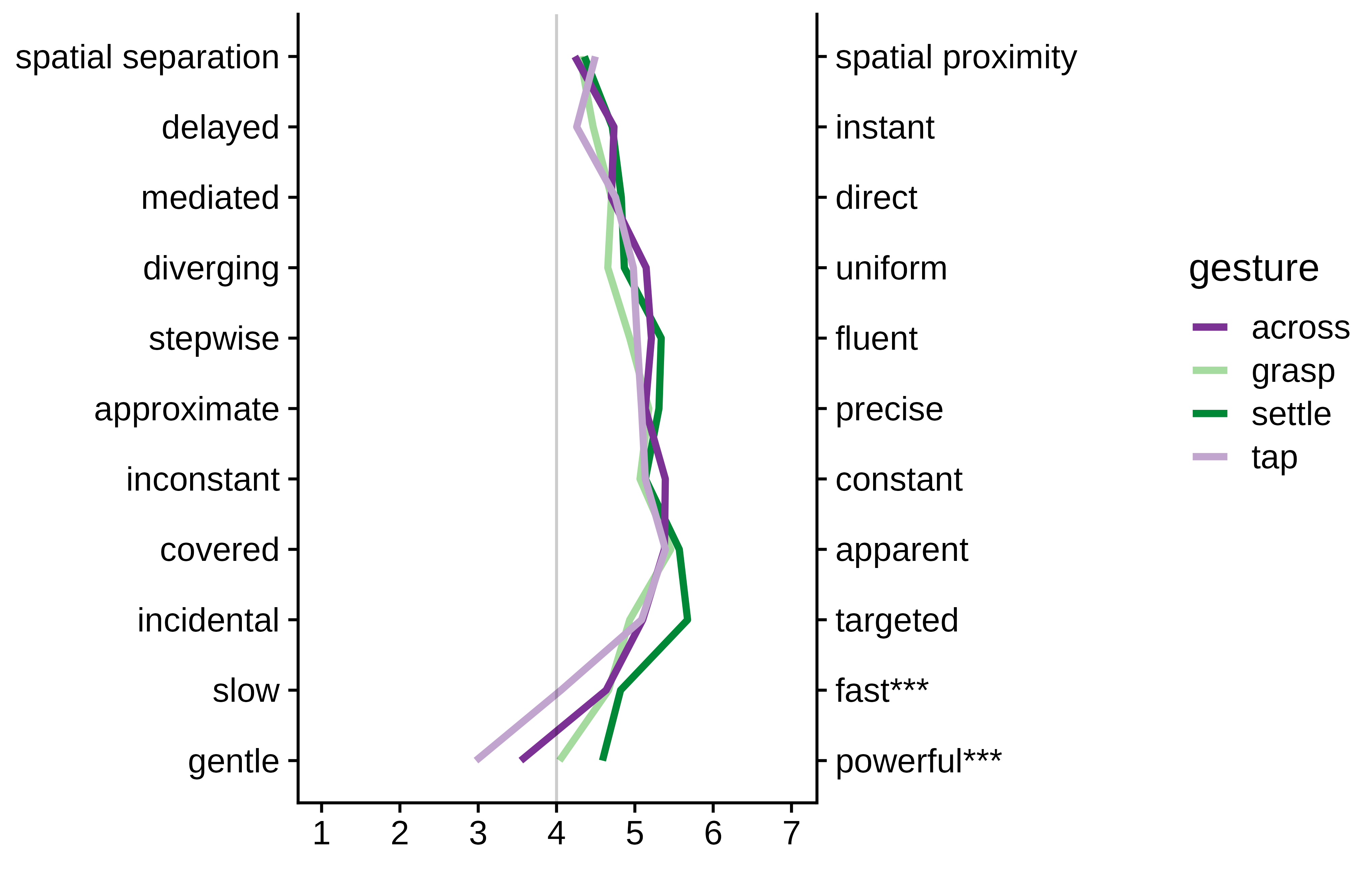}
\caption{Mean scores for all gestures on each item of the Interaction
  Vocabulary. ``Grasp'' and ``settle'' (green) make a
  spatial reference to the conversation partner. ``Across'' and ``settle''
  (darker tone) make a spatial reference to the source of disturbance. The
  gestures were significantly different in the attributes ``fast'' and
  ``powerful'' (p<.001).}%
\label{fig:iv-scores}
\Description{A graph that plots the mean scores for each gesture on each item of
  the Interaction Vocabulary. On most items, the four groups have relatively
  similar scores. The gestures were perceived as slightly more: instant (not delayed),
  direct (not mediated), uniform (not diverging), fluent (not stepwise),
  precise (not approximate), spatial proximity (not spatial separation),
  constant (not inconstant), and apparent (not covered). They were
  also perceived as targeted (not incidental), fast (not slow, except for the tap
  gesture), and powerful, not gentle (except for the tap and across gestures).}
\end{figure}

To that end, we first analyzed the overall witness perception according to the
form components of the Interaction Vocabulary (i.e., undetermined, energetic,
and straightforward). We computed the average of all participants' scores across
all gestures and compared it with the center of the scale (4.0) in an
undirected, one-sample t-test. Overall, participants perceived the gestures as
above-average energetic ($t(405)=3.94$, $p<.001$) and straightforward
($t(405)=23.13$, $p<.001$), and below-average ``undetermined'' ($t(405)=9.13$,
$p<.001$). Thus, some example attributes that describe the form of interaction
for all gestures are: Instant, direct, apparent, precise, uniform, fluent, and
constant (see Figure~\ref{fig:iv-scores}).

\subsubsection{``Settle'' more energetic than ``across'' and ``tap''; ``grasp''
more energetic than ``tap''}

Next, we looked for differences between the gestures. We first ran a one-way
ANOVA with gesture type as independent variable and the ``energetic'' component
as dependent variable and found a significant effect ($F(3, 402)=11.05$,
$p<.001$, $\eta^2_{p}=.08$). Follow-up, pairwise comparisons indicated that
witnesses perceived the ``settle'' gesture as more energetic than the ``across''
gesture ($t(192)=3.56$, $p_{Holm}<.01$) and more energetic than the ``tap''
gesture ($t(207)=5.37$, $p_{Holm}<.001$). In addition, they perceived ``grasp''
as more energetic than ``tap'' ($t(210)=3.83$, $p<.001$). No other differences
were significant.

We further explored these differences between gestures on the attribute level of
the ``energetic'' component. To that end, we ran four one-way ANOVAs, each with
gesture type as independent variable and one of the four attributes as dependent
variable (i.e., fast, powerful, instant, and direct).

\subsubsection{All other gestures faster than ``tap''}

This revealed a significant difference in the attribute ``fast'' ($F(3, 402)
= 5.23$, $p <.001$). Post-hoc, pairwise comparisons revealed that witnesses
perceived the ``settle'' gesture as faster than the ``tap'' gesture ($t(207)
= 3.69$, $p_{Holm} <.01$). ``Grasp'' was also perceived as faster than ``tap''
($t(210) = 2.96$, $p_{Holm} <.02$), and ``across'' was perceived as faster than
``tap'' ($t(193) = 2.96$, $p_{Holm} < .03$). No other differences were
significant.

\subsubsection{``Settle'' most powerful, followed by ``grasp'', then ``across'',
then ``tap''}

In addition, we found significant differences between all gestures on the
``powerful'' attribute ($F(3, 402) = 23.94$, $p < .001$). Post-hoc, pairwise
comparisons revealed that witnesses perceived ``settle'' as more powerful than
``grasp'' ($t(209) = 2.78$, $p_{Holm} <.02$), ``across'' ($t(192) = 5.04$,
$p_{Holm} < .001$), and ``tap'' ($t(207) = 8.12$, $p_{Holm} <.001$). They
perceived ``grasp'' as more powerful than ``across'' ($t(195) = 2.40$, $p_{Holm}
<.02$) and ``tap'' ($t(210) = 5.40$, $p_{Holm} <.001$). Finally, they perceived
``across'' as more powerful than ``tap'' ($t(193) = 2.78$, $p_{Holm} <.02$).

With two further one-way ANOVAs and gesture type as independent variable, we
tested for differences on the other two attributes, but found no significant
effects (``instant'' ($F(3, 402) = 2.13$, $p = .10$); ``direct'' ($F(3, 402)
= 0.20$, $p = .90$)). There were also no effects for the other two components of
the Interaction Vocabulary. A one-way ANOVA with gesture type as independent
variable and ``straightforward'' as dependent variable was not significant
($F(3, 402) = 2.20$, $p = .09$). Finally, a one-way ANOVA with gesture type as
independent variable and ``undetermined'' as dependent variable revealed no
significant effect ($F(3, 402) = 1.97$, $p = .12$).

All in all, the Interaction Vocabulary provides a more detailed picture of how
witnesses perceived the form of interaction of the four gestures. Overall, they
perceived the gestures as energetic and straightforward, but not as
undetermined. Among the individual gestures, ``settle'', which makes a spatial
reference both to the source of disturbance and the conversation partner, was
perceived as most energetic. It was perceived as faster than ``tap'' and more
powerful than all other gestures. Witnesses perceived ``tap'', which makes no
spatial references, as least energetic, slowest, and least powerful.

\subsection{Witness Experience}

We next analyzed the witness experience according to the three qualities
``secretive'', ``suspenseful'', and ``expressive'' (RQ3).

\subsubsection{Overall experiences: Secretive, expressive, but not suspenseful}

\begin{figure}[t]
\centering
\includegraphics[width=0.9\linewidth]{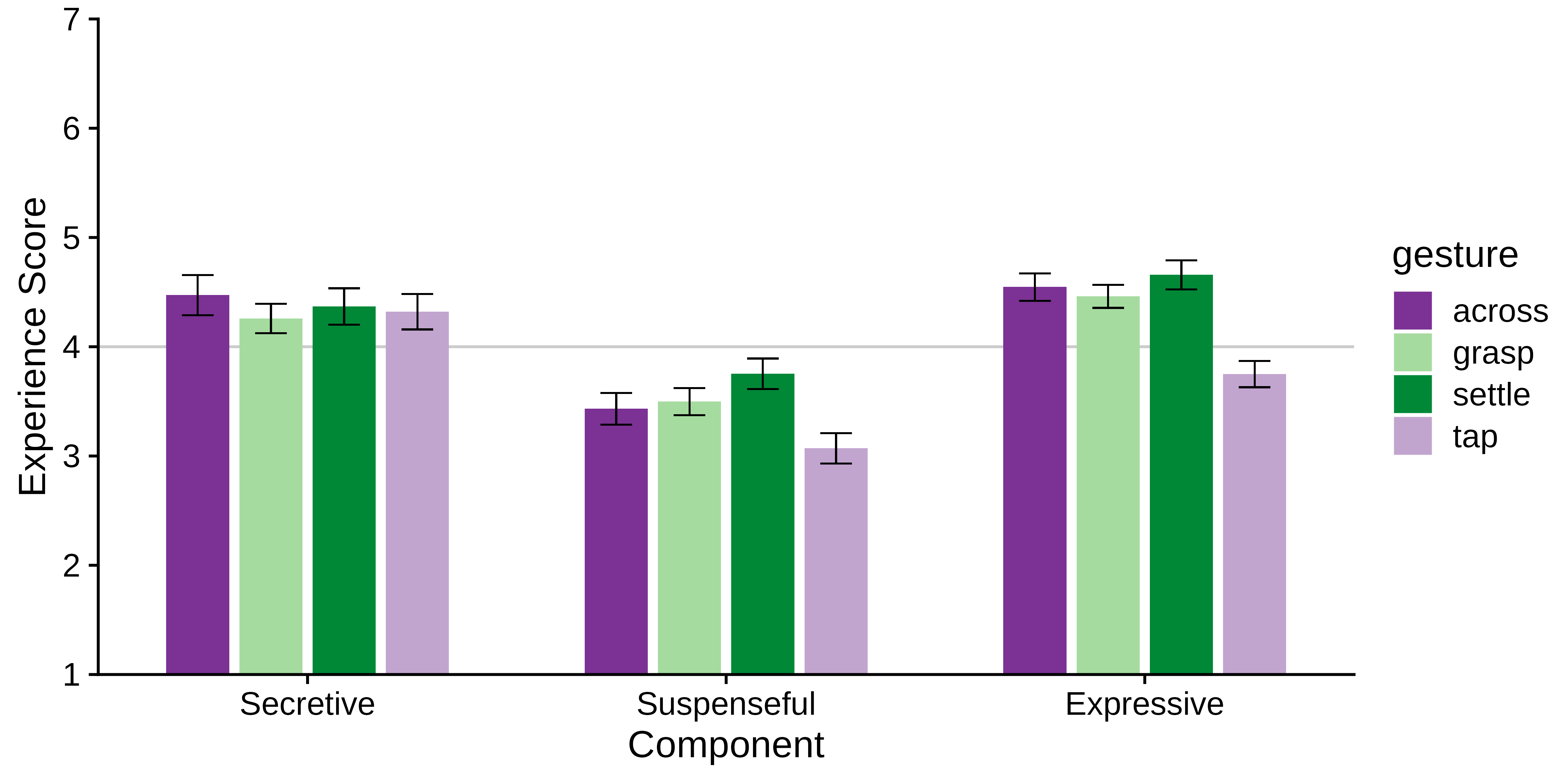}
\caption{Means and 95\%-confidence intervals for the four suspenseful
  interactions on the three experience components.
  ``Grasp'' and ``settle'' (green) make a
  spatial reference to the conversation partner. ``Across'' and ``settle''
  (darker tone) make a spatial reference to the source of disturbance.}%
\Description{This graph plots the means for each gesture on the three experience
  measures: Secretive, suspenseful, and expressive. All gestures are above the
  center of the scale for the secretive scale, and all but the tap gesture are
  above the center of the expressive scale. Conversely, all means are below the
  center of the scale for the suspenseful scale. This means that the gestures
  were experienced as somewhat secretive and expressive (except for ``tap''),
  but not suspenseful.}%
\label{fig:experience-barplot}
\end{figure}

Again, we first tested the overall experience across all four gestures by
summarizing the scores of all participants and comparing them with the center of
the scale (4) using an undirected, one-sample t-test. We found that participants
experienced the four gestures as above-average secretive ($t(406)=4.35$,
$p<.001$, $d=0.05$), and expressive ($t(406)=5.52$, $p<.001$, $d=0.05$), with
significantly higher scores than the center of the scale. In contrast, they
experienced them as less-than-average suspenseful ($t(406)=-8.11$, $p<.001$,
$d=-0.05$), with an overall score below the center of the scale (see
Figure~\ref{fig:experience-barplot}). This is surprising, because the gestures
were designed according to the suspenseful category from the taxonomy, and the
visibility profile also matched this category.

\subsubsection{``Settle'' more suspenseful than ``tap''}

We further investigated differences between the gestures for each experience
component. In the first one-way ANOVA with gesture type as independent variable
and suspensefulness as dependent variable, we found a significant difference
($F(3, 403)=4.36$, $p<.01$, $\eta^2_{p}=.03$). Post-hoc, pairwise comparisons
revealed that witnesses experienced ``settle'' as more suspenseful than ``tap''
($t(207)=3.58$, $p_{Holm}<.01$). No other pairwise comparisons were significant.

\subsubsection{``Tap'' less expressive than all other gestures}

The second, one-way ANOVA with gesture type as independent variable and
expressiveness as dependent variable also revealed a significant difference
($F(3, 403)=11.79$, $p<.001$, $\eta^2_{p}=.08$). Post-hoc, pairwise comparisons
showed that witnesses experienced ``tap'' as less expressive than ``across''
($t(194)=4.57$, $p_{Holm} < .001$), ``grasp'' ($t(210)=4.25$, $p_{Holm} <.001$),
and ``settle'' ($t(207)=5.39$, $p_{Holm}<.001$). No other pairwise differences
were significant.

Finally, the third one-way ANOVA with gesture type as independent variable and
secretive as dependent variable was not significant ($F(3, 403)=0.30$, $p=.83$).

In sum, witnesses experienced the gestures, overall, as both secretive and
expressive, but not as suspenseful. Among the gestures, we found that they
perceived ``settle'' as more suspenseful than ``tap''. They also perceived
``tap'' as less expressive than all other gestures.

\subsection{Combined Effects of Visibility and Form on the Witness Experience}

Lastly, we analyzed the impact of the perceived interaction on the witness
experience (RQ4) with a multiple regression analysis. Here, we combined a total
of seven predictors. These included the two visibility measures (manipulation
and effect) and the three form components (undetermined, energetic, and
straightforward). In addition, we included two further categorical predictors.
These predictors indicated whether the interaction makes a meaningful spatial
reference to the source (``across'' and ``settle'') and/or to the partner
(``grasp'' and ``settle''). Our goal with this analysis was to explore the
effect of visibility, form perception, and intended meaning of the form on
witness experience in a comprehensive analysis. We ran three analyses, with the
three experience components as outcome, respectively.

\subsubsection{Suspenseful visibility patterns create secretive experiences}

We ran the first regression analysis with the seven predictors and secretive
experience as outcome and found two significant predictor variables:
Manipulation visibility ($b=0.46$, $CI_{95\%}=[0.29, 0.62]$,
$r_{\text{part}}=.26$ $p<.001$) and effect visibility ($b=-0.28$,
$CI_{95\%}=[-0.44, -0.13]$, $r_{\text{part}}=-.17$ $p<.001$; $R^2=.11$). No
other predictor was significant. This confirms an effect of subjective
visibility on the secretive experience. However, the specific pattern is
interesting, and different from predictions according to the taxonomy.
Participants experienced the interaction as more secretive the \emph{more} they
perceived the manipulation as clearly visible, and the \emph{less} they
perceived the effect as visible. This visibility pattern is supposed to be
associated with suspenseful experiences, but led to secretive experiences in our
study.

\subsubsection{More effect visibility and spatial reference to partner create
suspenseful experiences}

The second regression had the same seven predictors and the suspenseful
experience measure as outcome. Here, we also found two significant predictors:
Effect visibility ($b=0.15$, $CI_{95\%}=[0.01, 0.29]$, $r_{\text{part}}=.10$
$p<.05$) and spatial reference to the partner ($b=0.35$, $CI_{95\%}=[0.07,
0.62]$, $r_{\text{part}}=.12$ $p<.05$; $R^2=.05$). Although significant, the
model fit in this regression was relatively low. Nevertheless, the findings
indicate two interesting tendencies. First, manipulation visibility was no
significant predictor of suspenseful experience ($p=.36$). Perhaps this was in
part a result of the high manipulation visibility of all gestures. Second, the
effect visibility was \emph{positively}, not negatively associated with
suspenseful experiences. This means that participants who perceived the effect
as more visible also experienced the interaction as more suspenseful. Thus,
effect visibility worked in the opposite direction than assumed in the taxonomy,
at least in the scenario tested in our study.

\subsubsection{Effect visibility, energetic form, and spatial relations all contribute to expressive experiences}

The third regression analysis used the same seven predictors, with the
expressive experience measure as outcome. Here, four predictors were
significant: Effect visibility ($b=0.36$, $CI_{95\%}=[0.25, 0.47]$,
$r_{\text{part}}=.28$ $p<.001$), energetic form ($b=0.32$, $CI_{95\%}=[0.20,
0.43]$, $r_{\text{part}}=.24$ $p<.001$), spatial relation to the source
($b=0.35$, $CI_{95\%}=[0.13, 0.57]$, $r_{\text{part}}=.14$ $p<.01$), and spatial
relation to the partner ($b=0.26$, $CI_{95\%}=[0.04, 0.49]$,
$r_{\text{part}}=.10$ $p<.05$; $R^2=.24$). In other words, predictors from all
three groups (visibility, perceived form, meaningful spatial relations)
contributed to expressive experiences. More visible effects, a more energetic
form, and spatial relations of the interaction to the source of disturbance and
to the partner led to stronger experiences of expressiveness.

\section{Discussion}

In this paper, we presented a large online study ($n=407$) to explore the
relationship between an interaction and how witnesses experience it. We compared
four ``suspenseful'' mid-air gesture-based interactions with a hearing aid and
their effects on subjective visibility, perceived form of interaction, and
subjective experience from the witness perspective.

In terms of pure visibility (RQ1), the witness perception of the interactions
matched the suspenseful interaction pattern of the taxonomy \citep{Reeves2005}.
This is in line with our selection of suspenseful interactions with revealed
(and thus highly visible) manipulations and hidden (thus less visible) effects.
The four different gestures had similar visibility patterns, and we found no
difference between them in terms of subjective visibility from the witness
perspective.

As far as the form of interaction is concerned (RQ2), witnesses perceived
interactions overall as ``energetic'' and ``straightforward'', but not as
``undetermined''. In terms of individual attributes, they perceived the four
gestures overall as targeted, apparent, precise, and fluent, among others.
However, we found differences between the gestures on the ``energetic''
component. The most energetic form was ``settle'', and ``tap'' was the least
energetic. These differences were reflected in perceptions of the ``slow --
fast'' and ``gentle -- powerful'' attributes. Witnesses perceived ``settle'' as
the fastest and ``tap'' as the slowest among the interactions. In addition, they
perceived ``settle'' as more powerful than ``grasp'', which was more powerful
than ``across'', which was more powerful than ``tap''.

In other words, although we found no difference between the gestures in terms of
pure visibility, an analysis of the more specific form of interaction revealed
existing differences that we could measure with the Interaction
Vocabulary~\citep{Diefenbach2013, Lenz2013}.

Next, we analyzed the witness experiences (RQ3) and found that participants
perceived all gestures as relatively secretive and expressive (except for
``tap''), but not as suspenseful. Among the gestures, they perceived ``settle''
as more suspenseful than ``tap''. In addition, participants perceived all other
gestures as more expressive than ``tap''. These findings are somewhat
surprising. The taxonomy suggests that gestures with visible manipulations and
less visible effects such as the ones studied here lead to suspenseful
experiences. In addition, secretive experiences are supposed to result from
invisible manipulations and effects, which is not how participants perceived the
gestures in our study. In other words, the experience patterns from our study do
not match predictions based on the taxonomy~\citep{Reeves2005}.

Finally, we explored how the visibility patterns, the form of interaction, and
spatial references of the gestures shape the witness experience (RQ4). We found
that higher perceived manipulation visibility and lower perceived effect
visibility predicted secretive experiences but not suspenseful experiences as
suggested in the taxonomy. Instead, suspensefulness was predicted by
\emph{higher} subjective effect visibility and spatial relations to the partner
(i.e., witness). In addition, higher effect visibility, energetic form of the
interaction, and spatial relations to the source of disturbance and the partner
(witness) predicted expressive experiences.

\subsection{Broader Impact}

The paper provides new insights into the relationship between interaction design
and witness experiences. We have started by identifying some problems with
central concepts used in the literature. Most prominently, the definitions of
``hiding'' and ``revealing'' are still too vague, and we present three different
understandings that are partially in conflict with each other. The taxonomy
mostly seems to refer to a perception-based understanding. This means that
``hiding'' parts of an interaction makes them sensorically invisible to
a witness, and ``revealing'' makes them visible. However, this understanding
seems to be in conflict with some of the taxonomy's own predictions.
Specifically, it suggests that secretive experiences result from interactions
the witness cannot even perceive. Here, the second understanding of hiding and
revealing, based on social conventions, comes into play. According to the
convention-based perspective, hidden interactions remove something that ``should
be there'', from the perspective of the witness, and revealed interactions make
something more prominent than usual. This perspective relates to previously
suggested ``subtle'' interaction design~\citep{Rico2010} and the notion of
``amplified'' interactions~\citep{Reeves2005}. Finally, according to the third
understanding, we can think of hiding and revealing as experienced by the
witness. Although this may relate to sensory perceptions and social conventions
as well, it changes the perspective from what the designer and user want to hide
or reveal to what the witness actually experiences as ``hidden'' and
``revealed''.

The study also indicates problems that can arise if interaction design continues
to overemphasize the user, while considering the witnesses and the broader
social situation only superficially (e.g.,~\citep{Kuutti2014, Baumer2017}).
A designer- or user-centered perspective is at risk to focus on the interaction
in isolation. But that perspective overlooks what witnesses can see and
interpret as effects. For example, all four gestures (manipulations) in our
study were clearly visible, but there was no effect of the technology to be
observed. Nevertheless, participants did not rate the effect as entirely
invisible. They ``saw'' something. Specifically, the interaction happened within
the frame of a conversation, which continued afterwards. From the witness
perspective, it can be difficult to clearly tell apart related or intended
effects from unrelated ones, because the interaction does not happen in
a neutral void. Even in an alternative scenario where the hearing aid does not
function as intended, witnesses might see the user struggling to make the
hearing aid work or indicate that they cannot hear them anymore. From the
witness perspective, these can also be understood as ``effects'' of the
interaction. A possible exception might be stage settings, as studied in the
original taxonomy~\citep{Reeves2005}. They define a clearly delimited, fully
observable subcontext within the broader social situation, in which the
performers may have more control of manipulations and effects they show to the
audience.

If visibility of manipulations and effects alone is insufficient to distinguish
meaningfully between different witness experiences, we need an alternative. In
the present study, we used the Interaction Vocabulary~\citep{Diefenbach2013,
Lenz2013} to more closely analyze effects of the form of interaction. Regarding
the taxonomy, we would suggest revising the distinction between ``visible'' and
``invisible'' parts of the interaction and use alternative descriptions. Reeves
and colleagues~\citep{Reeves2005} already presented visibility more as
a continuum, rather than binary categories. However, later work often fell back
to categories, presumably to fit gestures into experimental designs. We would
argue that there is more variation than pure visibility for designers to
explore. In our case based on mid-air gestures, for example, Hosseini and
colleagues~\citep{Hosseini2023} synthesized a set of gestures, based on
agreement rates of their meaning. These may not be strictly different in terms
of visibility. But they carry different situated meanings that could have
different effects on witnesses. Thus, the ``visibility'' axis of the taxonomy
should be revised. That said, we think that the taxonomy's distinction between
manipulation and effect design, and thinking about how the two go together,
still provides a solid framework for interaction designers.

Finally, a crucial factor for the witness experience that we have not covered
here and that is not covered in the taxonomy is what the witness is currently
doing, and how this relates to the user's interaction~\citep{Dourish2004,
Uhde2021b, Uhde2022d}. Uhde and Hassenzahl pointed out that even the same
interaction, with the same form of manipulations and effects, can create
different experiences for witnesses (see the phone call example in the
background section). The witness activity is challenging to anticipate in
dynamic social settings or for mobile technologies. We think that this
uncertainty may have led to the tendency towards ``defensive'' technology design
in the past (i.e., hiding most of the interaction;~\citep{Koelle2020}). Such
design patterns are relatively flexible and rarely conflict with other people's
activities. However, they come with the risk to be discovered and potentially
breach trust~\citep{Pohl2019}. In addition, hidden designs essentially remove
the technology from the social situation. In doing so, they give up on the
opportunity for technologies to create positive experiences, instead of merely
avoiding negative ones~\citep{Uhde2022d}.

\subsection{Limitations}

Of course, this study also has some limitations. First, we found the effects
only in the context of a cocktail party scenario and only based on interactions
with a hearing aid. We could argue that it still served the purpose to
demonstrate that the experience predictions made by the taxonomy cannot be
applied to social situations in general. However, we have not studied a broader
range of social situations and thus cannot qualify more precisely to which
extent and in which kinds of situations they still apply. The relationship
between interactions and experiences described in~\citep{Reeves2005} seems
plausible in the situations reported there.

The particular technology design may have also had an effect on witness
experiences. We used a futuristic hearing aid to avoid pre-existing mappings
from manipulations to effects that participants could already have with widely
used technologies. They may perceive the interactions differently with
smartphones or laptops. In addition, the form of the hearing aid was
unconventional. Although many hearing aids are currently designed to be subtle
or invisible, there are different opinions within the community of people with
divergent hearing about their design. Some people explicitly want more visible
designs and compare hearing aids with the established fashion status of glasses,
or they want to avoid misunderstandings (e.g., a hearing aid can communicate ``I
couldn't hear you'' rather than ``I don't understand you'';
\citep{Doerrenbaecher2019}). A more visible design may also normalize divergent
hearing and help reduce stigma over time. The hearing aid design from our
broader research project followed this more communicative approach. In our
particular scenario, independent of the form, the hearing aid had the additional
advantage to integrate meaningfully within the conversation scenario, without
introducing an unrelated technology with possible confounding effects.

A second limitation is our exclusive focus on suspenseful interactions. As
indicated above, we made this decision because we ultimately prioritized
variations in experience over variations in hidden and revealed manipulations
and effects, and this resulted in insightful findings. But experiences may also
differ in the expressive, magical, and secretive quadrants. A complementary,
detailed analysis of the path from interaction design to witness experience
based on all four categories from the taxonomy could add further insights.

Third, we outlined a series of steps that can happen between interaction design
and witness experience. The interaction designer selects a certain form of
interaction. Then the user performs that interaction. Then the witness perceives
that interaction, which finally leads to the witness experience. Between each of
these steps, some information may get lost or new, unrelated information may be
interpreted as related. We have only covered the witness perspective here, and
future work is needed to better understand the other steps.

\subsection{Design Implications}

\subsubsection{Consider ``hiding'' and ``revealing'' from the witness perspective}

We dedicated a longer section in the beginning of this paper on issues with the
concepts of hiding and revealing. When used as in the taxonomy (i.e., mostly
perception-based), hiding in particular does not always seem useful as a tool to
create a witness experience. As we pointed out, an interaction with a hidden
manipulation and effect probably creates no secretive experience, but rather no
experience at all. In contrast, when using the convention-based or
experience-based approaches to hiding and revealing, the relation to the
specific interaction design becomes more complex. Interactions may stand out
because they are less visible, or they imply through their form that something
is ``hidden'' or ``revealed''.

For example, we found that the interactions in our study led to more secretive
experiences if the participants perceived the manipulation as \emph{more}
visible. The visible manipulation is important to indicate that some effect can
be expected, but then nothing happens (or at least the witness is not involved).
We think that this can lead to interesting witness experiences, but it may be
difficult to design for. There is a fine line between successfully creating
interest or curiosity and ``trying too hard'' to catch someone's attention.
Goffman~\citep{Goffman1959} addressed a related topic and described scenarios of
how people try to communicate ``by accident'', for example that they are
popular. The crucial point in his framework is that the witnesses continue to
believe that the communication happens accidentally. Creating this illusion of
accidental communication can be an interesting area to explore for designers.

\subsubsection{Explore a richer repertoire of forms and aesthetics}

Even with a focus on the interaction alone, we found that hiding/revealing or
perceived visibility only accounted for parts of the witness experience. A more
detailed analysis of the form of interaction seems promising. We used the
Interaction Vocabulary \citep{Diefenbach2013, Lenz2013} here, which focuses on
aesthetic aspects and has already been applied in relation to user experiences
in the past \citep{Diefenbach2017c, Lenz2017}. Designers can use it as a first
extension of the hiding and revealing strategies. We also think that it could be
adapted to study manipulations and effects separately and combine, for example,
a fast manipulation with a slow effect or vice versa.

\subsubsection{Design for more defined social situations}

Social situations can be complex, and it seems ambitious for designers to
reliably shape the experience of people merely witnessing a user's performance,
simply by tuning the interaction's visibility patterns in a binary fashion. The
taxonomy that suggested such a relationship \citep{Reeves2005} was primarily
focused on stage settings, exhibitions, and other public performances with
a performer and an audience watching them. Although such settings can be full of
people, they often have some relatively predictable characteristics. Concert
visitors are positioned in a predictable location relative to the performer.
They come in typical clothing, and they perform certain predictable concert
activities, such as watching the stage performance. All of these characteristics
can be entirely different in other social situations, which has consequences for
how witnesses experience an interaction (see e.g., \citep{Uhde2021b}). Thus, we
can expect that suspenseful interactions (revealed manipulations and hidden
effects) will not reliably lead to suspenseful experiences across social
situations -- for example not in the conversation scenario we studied.

One ambition of today's technology design is to scale them to a broad set of use
settings. Mobile phones are the prime example for this, as they are used almost
everywhere. However, depending on the technology, designers can choose to take
a different approach and start by first focusing on one specific setting only.
For example, they could focus on a library, with its visitors and social norms,
which provides a relatively well-defined social situation. Designers can fit
their technology to such situations as much as possible, and only later extend
from there, rather than trying to address all possible settings at once. This
allows them to identify positive experiences that can result from their
interaction design in specific settings, and to keep them throughout the
technology's life cycle, rather than reducing the interaction to a minimal
acceptable version for all situations and thereby overlooking positive
opportunities.

\subsubsection{Consider witnesses as a source of positive experiences, not as
a problem}

Despite the complications indicated here and in previous studies, we encourage
designers to reconsider social situations, and think of them more as potential
sources of positive experiences and less as ``problems''. Social situations are
inherently associated with a loss of control. There are more people involved
than only the user, and everyone needs to adapt to a certain degree to the
others. People exchange some autonomy to be able to interact with other humans
(who do the same). Through our technology design, we can make a difference on
the nature of such social interactions. We can ``hide'' everything the user does
and maximize their ability to act independently of the others, neither
``disturbing'' them nor being disturbed. This is a ``safe'' route (with
exceptions, e.g.,~\citep{Uhde2022d}), but it only
leads to an acceptable or tolerable co-presence, not to enjoyable social
interactions. Conversely, interaction design that communicates more openly and
addresses others can make users more vulnerable to an extent. However, it also
allows for more positive social interactions and shared experiences. We see
a central role of designers as negotiators between these two perspectives, and
hope that future work more strongly considers the active, social involvement of
others.

\section{Conclusion}

In the present study, we outlined the long and mostly unexplored path from the
interaction design, to a user's performance, to a witness's perception of the
performance, and finally the witness experience. We shed some light on the
witnesses, and how they perceive and experience an interaction. The relationship
between interaction design and witness experiences turned out to be more nuanced
than previously framed. Although previous distinctions between hidden and
revealed manipulations and effects can be useful, HCI researchers and designers
should go beyond the current focus on mere visibility of an interaction, and try
to better understand the social situation in which an interaction is performed.
This can help identify ways to make a technology fit better with the situation
and create positive experiences. Finally, this could also make design
recommendations more relevant to guitar shop owners, who are still waiting for
better recommendations that keep their customers happy and themselves safe from
the ``forbidden riff''.

%%
%% The acknowledgments section is defined using the "acks" environment
%% (and NOT an unnumbered section). This ensures the proper
%% identification of the section in the article metadata, and the
%% consistent spelling of the heading.
\begin{acks}

This project is funded by the \grantsponsor{501100001659}{Deutsche
  Forschungsgemeinschaft (DFG, German Research
  Foundation)}{https://doi.org/10.13039/501100001659} -- Grant
  No.~\grantnum{425827565}{425827565} and is part
  of~\grantnum{427133456}{Priority Program SPP2199 Scalable Interaction
  Paradigms for Pervasive Computing Environments}. Parts of the research
  including the prototype were funded by the German Federal Ministry of
  Education and Research (Grant No.~\grantnum{16SV7786}{16SV7786}), through the
  mEEGaHStim project.

\end{acks}

%%
%% The next two lines define the bibliography style to be used, and
%% the bibliography file.
\bibliographystyle{ACM-Reference-Format}
\bibliography{bibliography.bib}

%%
%% If your work has an appendix, this is the place to put it.

\end{document}